\documentclass[11pt,a4paper]{article}
\usepackage[hyperref]{acl2021}
\usepackage{times}
\usepackage{latexsym}
\usepackage{mathtools}
\usepackage{graphicx}
\usepackage{bbm}
\usepackage{float}
\usepackage{array,multirow}
\usepackage[ruled,vlined]{algorithm2e}
\usepackage{algpseudocode}
\usepackage{booktabs}
\algnewcommand\algorithmicforeach{\textbf{for each}}
\algdef{S}[FOR]{ForEach}[1]{\algorithmicforeach\ #1\ \algorithmicdo}

\usepackage{microtype}

\aclfinalcopy 
\def\aclpaperid{1307} 

\newcommand{\yh}[1]{\textcolor{black}{{#1}}}
\newcommand{\zf}[1]{\textcolor{black}{{#1}}}

\title{A Query-Driven Topic Model}

\author{Zheng Fang$^1$, Yulan He$^1$$^,$$^2$ and Rob Procter$^1$$^,$$^2$ \\
  $^1$Department of Computer Science, University of Warwick\\
  $^2$The Alan Turing Institute, London, UK \\
  \texttt{\{Z.Fang.4|Yulan.He|Rob.Procter\}@warwick.ac.uk} }

\date{}

\begin{document}
\maketitle
\begin{abstract}

Topic modeling is an unsupervised method for revealing  the  hidden  semantic  structure  of  a corpus. It has been increasingly \yh{widely} adopted as a tool in the social sciences, including political science, digital humanities and sociological research in general. One \yh{desirable property of topic models} 
is to \yh{allow users to} find topics describing a specific aspect of the corpus. \yh{A possible} 
solution is to incorporate  domain-specific  knowledge into topic modeling, but this requires \yh{a specification from domain} experts. 
We propose a novel query-driven topic model that \yh{allows users to specify a simple query in words or phrases and return query-related topics, thus} avoiding \yh{tedious work from domain experts}. 
\yh{Our proposed approach is particularly attractive when the user-specified query has a low occurrence in a text corpus, making it difficult for traditional topic models built on word co-occurrence patterns to identify relevant topics.} Experimental results demonstrate the effectiveness of our model \yh{in comparison with both classical topic models and neural topic models}.
\end{abstract}

\section{Introduction}
Topic modeling \yh{aims to infer} 
topics from a collection of documents, where a topic is a salient pattern of the collection and is represented by a distribution over words. The availability in large volume of new sources of unstructured data, such as social media, has presented a challenge to conventional qualitative research methods in the social sciences and humanities and encouraged the exploration of topic modeling as a potential solution \cite{melville2019topic,hu2019hotel,yao2020tracking}. In these studies, topic modeling has been applied to questions centered on interpretation and meaning. By analyzing words distribution of topics learnt, researchers can apply inductive reasoning on specific topics and perform a more in-depth study of related documents, allowing them to identify underlying topical trends and conduct a more thorough analysis of the data.

One limitation of conventional topic modeling approaches in these studies is that they can only learn topics \yh{from} 
the whole corpus. However, in some cases, researchers may be interested in topics describing specific concepts or aspects of the corpus. To identify these topics, researchers have to analyze words distribution for all topics, thereby \yh{making it very time consuming}. 
Moreover, it could also happen that the target topics may have \yh{a very small} 
presence in the data to be detected directly by a topic model. For instance, given a set of posts about health, researchers may wish specifically to analyze the impact of food on health. If the words related to food have a relatively low frequency of occurrence in the posts, then conventional topic models such as \yh{Latent Dirichlet Allocation} (LDA) \cite{blei2003latent} may not find any \yh{food-related} topics 
at all. This is caused by the phenomenon of higher order co-occurrence in conventional topic models \cite{heinrich2009generic}, which prevents infrequent words being sampled under the correct topic. While 
an information retrieval method \yh{could be used} to find \yh{relevant} documents, \yh{identifying key subtopics discussed in these documents will still be a daunting process}. 

\begin{figure*}[htb] 
\centering 
\resizebox{0.96\textwidth}{!}{
\includegraphics[width=1\textwidth]{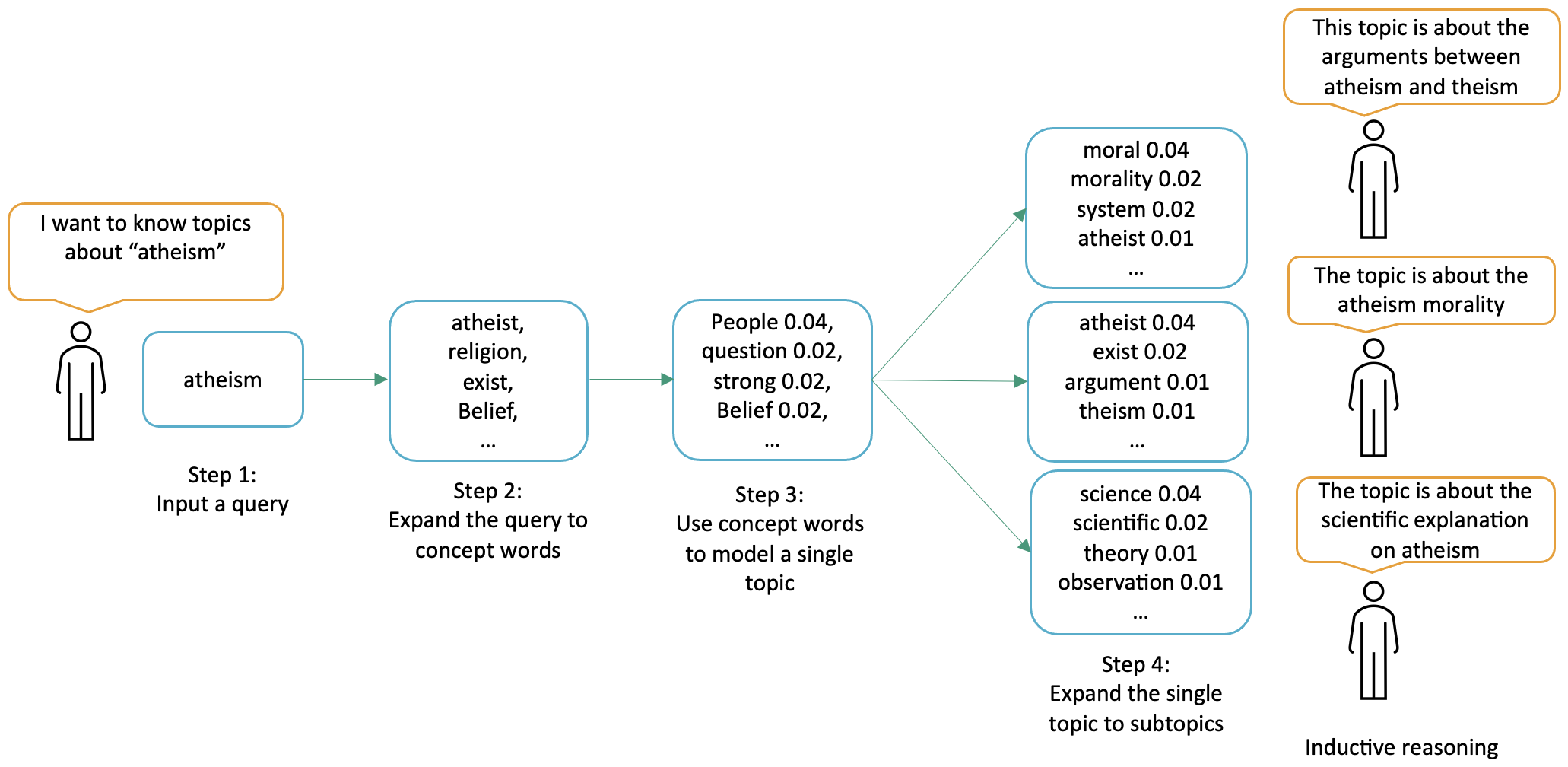} 
}
\caption{Our proposed model \yh{returns topics relevant to a user-input query, in this example, `\emph{atheism}'}. 
\textbf{Step 1}: user uses a query to define the concept of interest. \textbf{Step 2}: 
a query expansion technique \yh{is used} to expand the input query to a set of concept words. \textbf{Step 3}: the concept words \yh{are utilized} to generate a single topic. \textbf{Step 4}: the single topic is expanded to a set of subtopics. \yh{The retrieved concept-topic and subtopic results allow} 
the user to do inductive reasoning 
and have a more in-depth study of related documents. In Step 3 and Step 4, we present the top weighted words of the topic and \yh{their} corresponding weights. } 
\label{Fig.main2} 
\end{figure*}

To handle this limitation, weakly-supervised approaches \cite{andrzejewski2009latent,nikolenko2017topic,chen2013leveraging,andrzejewski2011framework,yang2015efficient} have been proposed as a solution and different types of domain-specific, prior knowledge, such as word correlation \cite{yang2015efficient}, document and word labels have been introduced. By adding these to the unsupervised topic model, a set of topics describing the domain knowledge can be generated. However, this still requires experts to define the domain knowledge, which may not always be feasible. In addition, \yh{the aforementioned approaches} 
can only generate one topic relevant to the target concept. It is \yh{desirable} 
to distinguish between different contexts about the same concept: for instance, for the concept `\emph{Middle East}', there \yh{might be subtopics relating to} 
Middle East conflicts and Middle East resorts, respectively. 
In our work, we propose a novel approach that 
automatically \yh{generates} 
all subtopics relevant to the target concept.

In our query-driven topic model, a query phrase is used to define the concept of interest. As illustrated in Figure 1, a query expansion technique is first employed to expand the input query to a set of concept words, which are then utilized to first generate a single topic about the concept, \yh{and subsequently} 
further expanded to a set of subtopics automatically. 
\yh{In summary, our contributions are four fold: 
\textbf{(1)} We propose a novel approach which} 
allows users without expertise knowledge to use a short query rather than predefined keywords to detect topics of their interests; 
\textbf{(2)} Our model is novel in its ability to identify rare topics in text, which would not be possible using existing topic modeling approaches;
\textbf{(3)} \yh{Our model is built on the Hierarchical Dirichlet Process (HDP) and can therefore} 
automatically infer all subtopics describing the target concept without having to determine the optimal number of topics 
beforehand; 
\textbf{(4)} \yh{We evaluate our approach on three datasets and achieve superior performance compared to both traditional hierarchical topic models and neural topic models, both quantitatively and qualitatively.}\footnote{Our source code can be accessed at: \url{https://github.com/Fitz-like-coding/QDTM}.} 

\section{Related Work}
Earlier work has attempted to solve the problem of identifying specific topics by using prior knowledge. \citet{andrzejewski2009incorporating} expressed domain knowledge with two primitives on word pairs called \emph{Must-Links} and \emph{Cannot-Links}, encoding them using a Dirichlet Forest prior. Topic-in-set knowledge \cite{andrzejewski2009latent} defines `\emph{z-labels}’ as prior knowledge and a similar idea was introduced by \newcite{nikolenko2017topic}. First-Order Logic has been proposed as a way to incorporate richer forms of prior knowledge \cite{andrzejewski2011framework}.  \citet{yang2015efficient} proposed an efficient method for incorporating domain knowledge and demonstrated significant speed improvement with large datasets. \citet{el2019semantic} presented a framework that allows users to incorporate the semantics of their domain knowledge in topic models interactively. \citet{gemp2019weakly} incorporated informative priors in an neural topic model for the purpose of semi-supervised topic modeling. All these approaches require experts to provide domain-specific, prior knowledge, which is problematic for two reasons: different corpora in the same domain may contain different information; and it may be costly to specify all prior knowledge. We take advantage of a query expansion technique and propose an automatic concept words extractor to help user extract prior knowledge. 

Our work is also similar to the Hierarchical Topic Model (HTM) \cite{blei2004hierarchical}. HTM is a non-parametric topic model that generates topics in a hierarchical structure. In our work, we also propose to generate subtopics from a parent topic. A key difference is that we propose a novel solution to incorporate domain-specific prior knowledge, making it possible to generate desirable topics. This is not the case with HTM. Although attempts were made to introduce prior knowledge in HTM, \citet{perotte2011hierarchically} focused on out-of-sample label prediction which is not the focus of our work while \citet{xu2018hierarchical} still required experts to define word pairs which is problematic as mentioned earlier.

\section{Proposed Framework}
In outline, our model expands an input query to a set of concept words using a concept words extractor. These concept words are then fed into a two phases framework based on a variant of a Hierarchical Dirichlet Process (HDP) to model all topics relevant to the concept. 

\subsection{Concept Words Extractor}
Given an input query $q$, we retrieve a list of documents $d$ according to the query likelihood score \cite{ceri2013introduction}, 
\begin{equation}\small
p(d{\mid}q) \approx \prod_{i=1}^{n}p(q_i{\mid}d)
\end{equation}
where $n$ is the number of tokens in the query and $p(q_i{\mid}d)$ is the probability of query term $q_i$ in document $d$. We define two extraction rules ``AND'' and ``OR'' to constrain whether query terms should appear in the same document or not. We then extract concept words from the retrieved documents. We adopt three approaches for our purpose.

\paragraph{Frequency based extraction (FRE)}
The first one simply extracts words with high frequency in the retrieved documents as our concept
words:
\begin{equation}\small
{Score(w)} = {\sum_{i}^{n}TF(w{\mid}d_i)} 
\end{equation}
where $n$ is the number of retrieved documents and $TF(w{\mid}d_i)$ is the term frequency of word $w$ in document $d_i$.

\paragraph{KL-Divergence based extraction (KLD)}
The second one is inspired by the query expansion technique \cite{carpineto2001information}. By intuition, words relevant to the input query have a high probability in the retrieved sub-corpus but a low probability in the whole corpus. The score can be defined as: 
\begin{equation}\small
{Score(w)} = {P_R (w)}log{\frac{P_R(w)}{P_C (w)}} 
\end{equation}
where $P_R (w)$ is the probability of word $w$ in the retrieved sub-corpus and $P_C (w)$ is the probability of word $w$ in the whole corpus. We extract words with high scores as our concept words. 

\paragraph{Relevance model with word embedding (REL)}
This approach extracts concept words from a word-embedding enhanced relevance model \cite{diaz2016query}. The probability assigned to word $w$ by the relevance model \cite{lavrenko2017relevance} is:
\begin{equation}\small
{p(w{\mid}RM)} = {\sum_{d{\in}R}p(w{\mid}d)p(d{\mid}q)} 
\end{equation}
where $R$ is the retrieved documents set, $p(w{\mid}d)$ is the probability of word $w$ in document $d$ and $p(d{\mid}q)$ is $d$’s query likelihood from equation (1). We integrate this model with word embeddings:
\begin{equation}\small
Score(w) = \lambda p(w{\mid}RM)+(1-\lambda)\mbox{sim}(w,q)
\end{equation}
where $\lambda$ is a hyperparameter and $\mbox{sim}(w,q)$ is the normalized similarity between word $w$ and the input query $q$. For each term in the vocabulary list, we calculate its similarity with the input query. We then take the top $k$ most similar terms and normalize their similarity values. If $w$ is among the top $k$ similar terms, $\mbox{sim}(w,q)$ would get the normalized similarity value. Otherwise, $\mbox{sim}(w,q)=0$. 



\subsection{Query-Driven Topic Model}

We propose a two-phase framework based on HDP, which is a nonparametric Bayesian model that can automatically infer the number of topics in a corpus \cite{teh2005sharing}. It assumes a restaurant (i.e., a document) has a set of tables and serves dishes (i.e., topics) from a global menu. A single dish is only served at a single table for all customers (i.e., words) who sit at that table. 


In the first phase, the model infers one topic for each concept, along with other irrelevant topics. We define this topic as the ``\emph{parent topic}" in later sections. We denote this parent topic of a concept corresponding to the input query $q$ as $\tilde{z}_q$. We incorporate prior knowledge into HDP by fixing the topic index for concept words in all documents. For words from concept words $W_q$ corresponding to the input query $q$, the topic index $z$ are known and remain fixed as $\tilde{z}_q$, and the probability for sampling an existing table $t$ for a word $w_{ji}$ at document $j$ and position $i$ in the Gibbs sampling process is: 
\begin{equation}\small
\resizebox{\columnwidth}{!}{$
  p(t_{ji}=t \mid t^{-ji},k) \propto \mathbbm{1_1}(w_{ji},k_{jt}) n_{jt}^{-ji} f_{k_{jt}}^{-w_{ji}} (w_{ji})
  $}
\end{equation}
where $k_{jt}$ is the topic assignment of table $t$ at document $j$ and  $f_{k_{jt}}^{-w_{ji}} (w_{ji})$ is the probability of $w_{ji}$ assigned to topic $k_{jt}$ after removing the current word and $\mathbbm{1_1}(w_{ji},k_{jt})$ is an indicator function, which takes on value 0 if $w_{ji} \in W_q$ and $k_{jt}{\neq}\tilde{z}_q$ and 1 otherwise. $n_{jt}^{-ji}$ denotes the number of words in document $j$ at table $t$ except the current word. The probability for sampling a new table $t^{new}$ is:
\begin{equation}\small
  p(t_{ji}=t^{new} \mid t^{-ji},k) \propto \alpha p(w_{ji} \mid t^{-ji},t^{new},{\bf k})
\end{equation}
where
\begin{equation}\small
\begin{split}
    p(w_{ji} \mid t^{-ji},t^{new},{\bf k})  = \sum_{k=1}^{k}\frac{m_ k}{m_\cdot+\gamma}f_k^{-w_{ji}} (w_{ji}) \\ +\frac{\gamma}{m_.+\gamma} f_{k^{new}}^{-w_{ji}} (w_{ji})
\end{split}
\end{equation}
Here, $m_k$ denotes the number of tables of topic $k$ and $m_\cdot$ denotes the total number of tables. $\gamma$ and $\alpha$ are the hyperparamenters of the model. $f_{k^{new}}^{-w_{ji}} (w_{ji})=\frac{1}{|V|}$ is the prior density of $w_{ji}$ where $|V|$ is the vocabulary size of the dataset. If the sampled table is a new table, we sample an existing topic $k_{j{t^{new}}}$ from:
\begin{equation}\small
  p(k_{j{t^{new}}} \mid t,k^{-j{t^{new}}})  \propto \mathbbm{1_1}(w_{ji},k_{jt})	{m_k} 	f_k^{-w_{ji}} (w_{ji})
\end{equation}
and probability for sampling a new topic $k^{new}$ is:
\begin{equation}\small
p(k_{j{t^{new}}}=k^{new} \mid t,{\bf k}^{-j{t^{new}}}) \propto \gamma f_{k^{new}}^{-w_{ji}} (w_{ji})
\end{equation}

In the second phase, the model expands the parent topic of each concept produced in the first phase to a set of subtopics. Let $W_{\tilde{z}_q}$ be the words assigned to the parent topic $\tilde{z}_q$ in the first phase, the probability for sampling an existing table $t$ for a word $w_{ji}$ in the Gibbs sampling process is:
\begin{equation}\small
  p(t_{ji}=t \mid t^{-ji},k) \propto \mathbbm{1_2}(w_{ji},k_{jt}) n_{jt}^{-ji} f_{k_{jt}}^{-w_{ji}} (w_{ji})
\end{equation}
where $\mathbbm{1_2}(w_{ji},k_{jt})$ is an indicator function that takes on value 1 if $w_{ji}{\in}W_{\tilde{z}_q}$ and $k_{jt}=\tilde{z}_q$ and 0 otherwise. Probability for sampling a new table $t^{new}$ is:
\begin{equation}\small
\begin{split}
 p(t_{ji}=t^{new} \mid t^{-ji},k) \propto \\
 \mathbbm{1_2}(w_{ji},k_{jt}) \alpha p(w_{ji} \mid t^{-ji},t^{new},{\bf k})
\end{split}
\end{equation}
where 
\begin{equation}\small
\begin{split}
  p(w_{ji} \mid t^{-ji},t^{new},{\bf k}) = \sum_{k=1}^{K}\frac{m_ k}{m_.+\gamma}f_{k}^{-w_{ji}} (w_{ji}) \\
  +\frac{\gamma}{m_.+\gamma} \hat{f}_{k^{new}}^{-w_{ji}} (w_{ji})
 \end{split}
\end{equation}
and $\hat{f}_{k^{new}}^{-w_{ji}} (w_{ji})=\frac{1}{|W_{\tilde{z}_q}|}$ is the prior density of $W_{\tilde{z}_q}$ where $|W_{\tilde{z}_q}|$ is the vocabulary size of $W_{\tilde{z}_q}$. For a new table, probability for sampling an existing topic $k$ is:
\begin{equation}\small
\begin{split}
p(k_{j{t^{new}}}=k \mid t,{\bf k}^{-j{t^{new}}}) \propto \\ \mathbbm{1_2}(w_{ji},k_{jt})	{m_k} 	{f_{k}}^{-w_{ji}} (w_{ji})
\end{split}
\end{equation}
and probability for sampling a new topic $k^{new}$ subordinate to the parent topic $\tilde{z}_q$ is:
\begin{equation}\small
p(k_{j{t^{new}}}=k^{new} \mid t,{\bf k}^{-j{t^{new}}}) \propto \gamma \hat{f}_{k^{new}}^{-w_{ji}} (w_{ji})
\end{equation}
The model automatically decides the number of subtopics and
we treat the subtopics produced as the final topics relevant to the target concept.

\paragraph{Incorporating Generalized P\'olya Urn scheme}
To make topics more interpretable, we incorporate word-embeddings by the Generalized P\'olya Urn scheme \cite{li2016topic}. P\'olya Urn scheme is introduced for colored balls and urns. In the Generalized P\'olya Urn scheme, when we draw a ball of a particular color, two balls of the same color are put back along with a certain number of balls of the similar colors. In topic modeling context, a topic can be viewed as an urn while a word can be viewed as a ball in a certain color and its semantically related words can be viewed as balls of similar colors. Every time we sample a word $w$ under a parent topic $\tilde{z}_q$, we increase the probability of sampling $w$ under $\tilde{z}_q$, as well as its semantically related concept words. Given pre-trained word embeddings, we calculate the cosine similarity between word $w_i$ and concept word $w_q\in W_q$. 
We then construct a word semantic relatedness matrix $\mathcal{M}$  \cite{li2016topic}, consisting of all word pairs whose cosine similarity is greater than a predefined threshold. We then construct a promotion matrix $A$ whose elements are efined as: 
\begin{align}\small
    & A_{i,q} =
    \begin{cases}
  				 1,	& \text {if $(w_i,w_q){\in}\mathcal{M}$ and $w_i = w_q$} \\
                 u,	& \text {if $(w_i,w_q){\in}\mathcal{M}$ and $w_i \neq w_q$} \\
                 0,	& \text {otherwise}
    \end{cases}
\end{align}
where $u{\in}(0,1)$ is a predefined promotion weight. When we sample a word $w$ under topic $\tilde{z}_q$, we also promote all its semantically related concepts words based on the amount of promotion in $A$. 

\paragraph{Word filtering} Inspired by \citet{wang2020optimising}, we propose a word filtering strategy. Word filtering can be used to prevent words that have weak ties with the sampled topic being promoted. For a word $w$ at $i^{th}$ Gibbs sampling iteration, its semantic cohesion to topic $k$ is:
	\begin{equation}\small
	CV[k,w_i]=\sum_{m=1}^{M}p^i(k,m){\cdot}\mbox{cos}\big(w_i,RW^i (k,m)\big)
	\end{equation}
where $p^i(k,m)$ is the probability of $m^{th}$ representative word in topic $k$ at $i^{th}$ iteration and $M$ is the number of representative words predefined. The representative words of topic $k{\neq}\tilde{z}_q$ at $i^{th}$ Gibbs sampling iteration are defined by the words ranked by the topic-word probability in the descending order. $\mbox{cos}\big(w_i,RW^i (k,m)\big)$ is the cosine similarity between word $w_i$ and the $m^{th}$ representative word of topic $k$ at $i^{th}$ iteration. The representative words of $\tilde{z}_q$ are simply its concept words. 

For the semantic cohesion of word $w$ with different topics $CV[\cdot,w]$, we map $CV[\cdot,w]$ into an arithmetic progression $\tilde{CV}[\cdot,w]$ ranging from 0 to 1.0 \cite{wang2020optimising}. We use the following equation to decide if the GPU is applied to $w$:
\begin{align}
\begin{split}
	S_{j,w}{\sim}Bernoulli(\lambda_{w,k_w}) \\
	\lambda_{w,k} = \frac{\tilde{CV}[k,w]}{\tilde{CV}_{max}[k,w]}
\end{split}
\end{align}
where $S_{j,w}$ indicates whether GPU is applied to word $w$ given document $j$ and ${\tilde{CV}_{max}}[k,w]$ is the maximal semantic cohesion among all topics.

We present the details of the Gibbs sampling process of the first phase of our model in Algorithm 1. We omit the details of the second phase of our model since it is similar to the first phase. The details of the functions Initialize($\cdot$) and UpdateCounter($\cdot$) can be found in Appendix C.

\begin{algorithm}[htb]
\small
\SetKwInput{KwInput}{Input}                
\SetKwInput{KwOutput}{Output}              
\SetAlgoLined
\KwInput{initial topic number $K$, hyperparameters $\alpha, \beta, \gamma$, word semantic relatedness matrix $\mathcal{M}$, documents $D$, and Concept words $W_Q$, }
\KwOutput{The posterior topic-word distribution}

Initialize($K$, $D$, $W_Q$)\; 
/* first phase */\;
\ForEach {$iteration$}{
    Update word-topic coherence using Eq.17\\
    \ForEach {document $j \in D $}{
        \ForEach {position $i \in j $}{
            Assign table $t \gets t_{ji}$\\
            Assign topic $k \gets k_{jt}$\\
            UpdateCounter($S_{j,w_{ji}}$, $t$, $k$, $False$)\;
            $t_{ji} \gets t \sim p(t_{ji}=t \mid t^{-ji},k)$ (Eq.6-8)\\
            \If{$t == t^{new}$}{
                $k_{jt} \gets k \sim p(k_{j{t^{new}}}=k \mid t,k^{-j{t^{new}}})$ (Eq.9-10)\\
            }
            $S_{j,w_{ji}} \gets updateGPUFlag(j,w_{ji})$ based on Eq.18\\
            UpdateCounter($S_{j,w_{ji}}$, $t_{ji}$, $k_{jt}$, $True$)
        }
    }
}
 \caption{Query-driven topic model.}
\end{algorithm}

\section{Experiments}
We conducted our experiments by two steps. In the fist step, we evaluated the quality of the parent topics from the first phase of our model. In the second step, we evaluated the quality of the subtopics  from the second phase of our model.

\subsection{Setup}
\paragraph{Datasets} We conducted our experiments on three datasets: 20Newsgroup\footnote{http://qwone.com/~jason/20Newsgroups/} contains around 18k newsgroup posts on 20 topics; TagMyNews\footnote{http://acube.di.unipi.it/tmn-dataset/} contains around 32k short English news from 7 categories; SearchSnippets \cite{xu2017self}  contains 12k short web search snippets from 8 categories. 

\paragraph{Baselines} We compared our model with six baselines: LDA \cite{blei2003latent} is a widely used topic model; DF-LDA \cite{andrzejewski2009incorporating} incorporates domain knowledge in LDA with Must-Links and Cannot-Links; SCLDA \cite{yang2015efficient} expresses prior knowledge as sparse constraints; ISLDA \cite{nikolenko2017topic} fixes topic index $z$ for certain keywords in all documents. AVITM \cite{srivastava2017autoencoding} is a neural topic model based on autoencoding Variational Inference. We also compared our model with BERT \cite{devlin2018bert}, a well-known neural language model, to test its document retrieval ability. To evaluate the quality of our subtopics towards the target concepts, we compared our model with HTM \cite{blei2004hierarchical}, which is also a nonparametric Bayesian model that can generate subtopics from higher level topics. 

\paragraph{Parameterization}
We set $\alpha=0.5$, $\beta=0.1$ for DF-LDA and $\alpha=0.1$, $\beta=0.01$ for SCLDA as suggested by the original papers. We set $\alpha=1/K$, $\beta=1/K$ for LDA and ISLDA, where $K$ is the number of topics pre-set for the models, and found it outperforms the original settings. We set $\alpha=0.1$, $\gamma=0.1$ and $\eta=0.01$ for HTM as suggested by the original paper and set the topic hierarchy depth to 3, to make it easier to compare with our model since topics from the second level of HTM can be considered as the parent topics and those from the third level as the subtopics of the parent topics. We set $\alpha=1.0$, $\beta=0.5$ and $\gamma=1.5$ as in the original HDP paper for our query-driven model, and  set the threshold for the cosine similarity used for the Generalized P\'olya Urn scheme to 0.5 and the promotion weight $u$ to 0.3. The number of representative words $M$ for the word filtering strategy was set to 10. $\lambda$ for the REL query expansion technique was set to 0.5 and $k$ was set to 100. 
In our experiments, we treated each category as a concept and determined the number of topics for each baseline model based on the number of categories in the datasets. For example, if a dataset had 16 categories, we set the number of topics to 17, using an extra one representing irrelevant information. 

\begin{table*}[htb]
\centering
\resizebox{\textwidth}{!}{
\begin{tabular}{lccccccccc}
\toprule 
\bf Model & \multicolumn{3}{c}{20news} & \multicolumn{3}{c}{TagMyNews} & \multicolumn{3}{c}{SearchSnippets}  \\ 
\cmidrule(lr){2-4} \cmidrule(lr){5-7} \cmidrule(lr){8-10}
 & Acc & coherence & Precision@K & Acc & coherence & Precision@K & Acc & coherence & Precision@K \\  \midrule
LDA & 0.650 & 0.420 & 0.588  & 0.781 & 0.384 & 0.687 & 0.804 & 0.390 & 0.696 \\
DFLDA & 0.623 & 0.421 & 0.562 & 0.772 & 0.386 & 0.649 & 0.795 & 0.390 & 0.644 \\
SCLDA & 0.666 & 0.402 & 0.622 & 0.804 & 0.418 & 0.745 & 0.816 & 0.414 & 0.796 \\
ISLDA & 0.680 & 0.406 & 0.645 & 0.801 & 0.411 & 0.729 & 0.845 & 0.421 & 0.804 \\
BERT & $-$ & $-$  & 0.156 & $-$  & $-$  & 0.300 & $-$  & $-$  & 0.261 \\
AVITM & 0.504 & {\bf 0.494} & 0.381 & 0.728 & 0.482 & 0.583 & 0.678 & 0.461 & 0.596  \\
\midrule 
Query-driven model(FRE) & {\bf 0.707} &  0.433 & {\bf 0.679} & 0.807 & 0.429 & 0.716 & 0.858 & 0.469 & 0.807 \\
Query-driven model(REL) & 0.601 & 0.452 &  0.557 & {\bf 0.837} & 0.444 & 0.749 & 0.851 & 0.501 & 0.780 \\
Query-driven model(KLD) & 0.705 &  0.435 &  0.677 & {\bf 0.828} & 0.414 & {\bf 0.755} & 0.860 & 0.465 & 0.811 \\
$+$ expert keywords & 0.690 & 0.430 & 0.659 & 0.817 & {\bf0.493} & 0.747 & {\bf 0.864} & {\bf 0.503} & {\bf 0.841} \\
\midrule 
$-$ word filtering & 0.706 & 0.434 & 0.678 & 0.823 & 0.413 & 0.744 & 0.859 & 0.432 & 0.811 \\
$-$ GPU & 0.698 & 0.400 & 0.671 & 0.818 & 0.389 & 0.730 & 0.849 & 0.413 & 0.808 \\
\bottomrule
\end{tabular}}
\caption{\label{font-table} Parent topic evaluation results for three datasets. FRE indicates frequency based query expansion; REL indicates Word embedding enhanced query expansion; KLD indicates  KL-Divergence based query expansion. BERT here is only for document retrieval purpose therefore we can't present the accuracy and coherence in the table.}
\end{table*}

For all baseline models, we asked an expert to provide prior knowledge. Each category in a dataset was associated with 10 keywords provided by the expert. For DF-LDA, we converted keywords to must-links. 
Since LDA, DF-LDA, AVITM and HTM cannot reveal the relationship between a concept and the generated topics directly, we need a further step to find the relationship between them. We calculated the average pairwise cosine similarity of the keywords and the top-10 word embeddings of each topic, and chose the topic with the highest similarity as the target topic of the concept. For HTM, we use the topics from the second level of its generated topic hierarchy. For our model, we used query phrases to represent the main concept of each category. Query phrases were interpreted directly from category names, e.g., we used ``\emph{computer graphics}'' to represent the category ``\texttt{comp.graphics}'' in the 20Newsgroup dataset. We removed categories that do not have meaningful names due to the difficulty of defining the query phrases for these categories, e.g., ``\texttt{talk.politics.misc}'' in the 20Newsgroup dataset. We then selected the top 10 concept words of each query based on the scores from the concept words extractor. We list expert-defined keywords and query phrase for each category in Appendix A. All models were trained until convergence. For BERT, we simply used the query phrases to retrieve relevant documents. We ran each model five times and present their average performance. 

\subsection{Parent Topic Evaluation} 

We evaluated the quality of parent topics of our model in terms of document classification, topic coherence and document retrieval performance. For document classification, a logistic regression classifier with default parameter settings was used. We used the topic distribution of each document as the input and conducted five-fold cross-validation. The topic distribution of a document represents the probability of each topic in a document. The quality of the topics can be assessed by the accuracy of text classification using the topic-level representation. A better classification accuracy means better latent semantic representations of the topics, indicating the learnt topics are more discriminative and representative. For topic coherence measure, we followed Roder et al. \shortcite{roder2015exploring} and used the best performing topic coherence measure C\_V based on the external corpus (Wikipedia). We focused on the top 10 words of our parent topics and used the Palmetto library algorithm \cite{roder2015exploring}. Higher coherence indicates better topic interpretability. For document retrieval, we adopted the metric ``precision@K" (P@K), which corresponds to the number of relevant results among the top $K$ documents. We retrieved documents of each topic based on the probability of the topic in the documents $p(z|d)$. If a topic can describe the target concept well, then the top retrieved documents should be relevant to the concept. In our experiments, we know the ground-truth number of documents belonging to each category, therefore we set $K$ for each concept to the actual number of documents from the corresponding category. We only considered the parent topics and reported the average results. A higher score indicates the model retrieves more concept-relevant documents, which is important when a researcher wants to do a more in-depth study of related documents.

Table 1 shows the performance of our models using three different query expansion techniques as well as only using expert-defined key words as prior knowledge. It can be observed that our models using the FRE and KLD query expansion techniques outperform all baselines except AVITM on almost all measures, though our model using the FRE query expansion technique has slightly worse document retrieval performance on the TagMyNews dataset. Although the model using REL is not as competitive as the models using FRE and KLD on the 20newsgroup dataset, it has the highest coherence scores on all the datasets, despite not using any expert-defined keywords as prior knowledge. This shows that combining word embeddings in query expansion can help produce more coherent prior knowledge. Although AVITM has better coherence score than our model on the 20newsgroup and TagMyNews datasets, its poor document retrieval performance indicates it is unable to find documents relevant to the target concept. Comparing our query-driven model with or without using expert-defined keywords, it achieves better coherence scores on the 20Newsgroup dataset without expert-defined keywords, though it performs slightly worse on the other two datasets. 

As for document classification and document retrieval, our KLD-based model has better performance than expert-defined keywords on 20newsgroup and TagMyNews, but does not work well on SearchSnippets.  This may be because our concept words extractor does not work well on short texts. The concept words extracted from TagMyNews and SearchSnippets are not as competitive as expert defined keywords. In addition, different interpretations of the same concept word may also compromise the performance. Interestingly, we also observed that BERT does not work well on these datasets. Possibly this is because we are using short query phrases to represent the concepts and BERT only works well for long queries. A short query may not give enough information about the concept that's why we adopted query expansion and topic modeling approaches.  Table 2 shows example concept-specific topics extracted from the TagMyNews dataset. It can be seen that extracted topics are closely related to their respective concept phrase. Topic extraction results on the other two datasets are shown in Appendix B.

\begin{table}[htb]
\centering
\resizebox{0.48\textwidth}{!}{
\begin{tabular}{lp{6cm}}
\toprule
{\bf Concept Phrase} & {\bf Top 10 words} \\
\midrule
{\bf business} & profit, business, bank, sell, usa, sale, credit, price, billion, stock \\
\midrule 
{\bf entertainment} & week, show, star, theater, pop, time, film, tv, sony, wedding \\
\midrule 
{\bf health} & disease, health, study, care, risk, drug, cancer, insurance, people, usa \\
\midrule 
{\bf technology} & google, apple, technology, ipad, china, company, online, intel, service, network \\
\midrule 
{\bf sport game} & game, play, playoff, final, boston, win, series, scored, season, sport \\
\bottomrule
\end{tabular}}
\caption{\label{font-table} Target concept topics for TagMyNews dataset.}
\end{table}

\begin{table*}[htb]
\centering
\resizebox{0.96\textwidth}{!}{
\begin{tabular}{lccccccccc}
\toprule
\bf Model & \multicolumn{3}{c}{20news} & \multicolumn{3}{c}{TagMyNews} & \multicolumn{3}{c}{SearchSnippets}  \\ 
\cmidrule(lr){2-4} \cmidrule(lr){5-7} \cmidrule(lr){8-10}
 & Diversity & Cohesion & overall & Diversity & Cohesion & overall & Diversity & Cohesion & overall\\  \midrule
{\bf HTM } & {\bf 0.94} & 0.54 & 0.51 & {\bf 0.93} & 0.53 & 0.49 & {\bf 0.86} & 0.49 & 0.42  \\
{\bf Query-driven model  } & 0.71 & {\bf 0.79} & {\bf 0.56} & 0.68 & {\bf 0.79} & {\bf 0.54} & 0.74 & {\bf 0.76} & {\bf 0.56}  \\
\bottomrule
\end{tabular}}
\caption{\label{font-table} Topic diversity and cohesion results for subtopics.}
\end{table*}

\paragraph{Ablation study:} We also studied the effectiveness of two major components in the proposed model: 1) GPU to incorporate word embeddings; 2) word filtering to remove unimportant words. The last two rows in Table 1 show the performance of our model using KLD query expansion technique without GPU and word filtering components. These show that GPU has a big impact for coherence and can help improve other measures in some extents, while removing word filtering reduces performance on all measures.

\begin{table*}[htb]
\centering
\resizebox{0.96\textwidth}{!}{
\begin{tabular}{lp{6cm}p{6cm}}
\toprule 
 {\bf Concept phrase} & {\bf Parent topic} & {\bf Subtopics }\\
\midrule
{\bf atheism} & {people, question, strong, belief, make, god, thing, point, religion, lack} & moral, system, morality, society, nature, objective, human, dream, animal, action\\\cmidrule(lr){3-3}
&  & people, question, thing, make, point, god, argument, belief, claim, true\\\cmidrule(lr){3-3}
& & science, scientific, theory, result, observation, scientist, experiment, hypothesis, methodology, model\\
\midrule
{\bf for sale} & {sell, sale, original, sold, interested, included, price, offer, box, cd} & sale, offer, price, sell, original, shipping, box, condition, interested, cd\\\cmidrule(lr){3-3}
 &  & list, interested, send, mail, address, post, email, original, information, call\\
 \midrule
 {\bf business} & {3do, government, key, phone, technology, company, business, chip, encryption, clipper} & company, phone, 3do, technology, business, number, system, japanese, computer, make\\\cmidrule(lr){3-3}
 &  & government, clipper, encryption, chip, system, nsa, phone, people, security, key\\\cmidrule(lr){3-3}
 &  & key, chip, algorithm, number, clipper, encryption, de, escrow, system, secret\\
\bottomrule
\end{tabular}}
\caption{\label{font-table} Parent topic and the subtopics of the concepts ``\emph{atheism}", ``\emph{for sale}" and ``\emph{business}" for the 20newsgroup dataset.}
\end{table*}

\subsection{Subtopics Evaluation}
We used our model with the KLD-query expansion technique in this evaluation. We dropped subtopics that have prevalence of less than $0.5\%$ in the corpus as these subtopics usually are not of interest. We evaluated the quality of subtopics in terms of topic diversity and topic cohesion\footnote{Note that \emph{topic cohesion} is different from \emph{topic coherence} as \emph{topic cohesion} measure the relevant between a subtopic and its parent topic.}. Topic diversity measures how much a subtopic overlaps with each other. We define it to be the percentage of the unique words in the top 25 words of all subtopics subordinate to the same parent topic \cite{dieng2020topic}. Higher diversity indicates more varied topics, while lower diversity indicates more redundant topics. Topic cohesion measures the relevance between the subtopics and the parent topic. We define it to be the cosine similarity between the parent topic embedding and the subtopic embedding. We can get the topic embedding as the weighted summing of the embeddings of its top 10 associated word. We combine these two metrics and define the overall quality of a subtopic as their product.

We report the results in Table 3. It shows that our model outperforms HTM by topic cohesion measures on all datasets, though with lower topic diversity scores. The high cohesion score indicates that the subtopics of our model is highly relevant to the target concept. By taking both measures into a account, our model achieves relatively better performance. It is expected since we  incorporate domain prior knowledge into our model. 

\subsection{Qualitative Evaluation}

We present the qualitative evaluation results in this section. We show a set of topics produced by our model from the 20newsgroup dataset in Table 4. The input concept phrases are shown in the left side of the table. For the concept phrase ``\emph{atheism}", which means the absence of belief in the existence of deities, we can see that our parent topic is highly relevant to it. The topic words like ``\emph{question}", ``\emph{belief}", ``\emph{god}" and ``\emph{lack}" clearly indicate that the topic is related to the arguments about God. By looking at the subtopics, we can easily see the first subtopic is about the atheism morality, 
the second subtopic is about the arguments between atheism and theism, 
and the third subtopic is about the scientific explanation on atheism. 
For the concept phrase ``\emph{for sale}", which means selling an item in a cheaper price, our parent topic includes many relevant words, such as ``\emph{sell}", ``\emph{sale}", ``\emph{sold}" and ``\emph{price}". The inclusion of ``\emph{box}" is less easy to explain, but could be related to product packaging. As expected, our subtopic reveals a sub-aspect about the concept that can not be identified directly from the parent topic: the email subscription for the latest news. This is reasonable, since merchants usually use email to provide customers with information about the latest products. The words like ``\emph{interested}", ``\emph{mail}", ``\emph{send}", ``\emph{information}" and ``\emph{original}" provide more information about the concept. 

\zf{The last row of Table 4 presents the topics of the low occurrence query ``\emph{business}", appeared only 294 times in the corpus, which is extremely low compared with the majority of other words in the corpus. LDA and HTM are unable to generate relevant topics due to the aforementioned ``higher order co-occurrence" issue, but our model can produce reasonable topics. The topic words ``\emph{encryption}", ``\emph{key}", ``\emph{phone}", ``\emph{company}" and ``\emph{business}" in the parent topic shows that the topic is related to data encryption for business. By looking at the subtopics of the topic, we can get a rough idea that the first subtopic is about a  Japanese phone company, since the top weighted words include ``\emph{japanese}", ``\emph{company}", ``\emph{phone}", ``\emph{technology}". This makes sense since phone companies usually have a strong requirement for encryption. The second and third subtopics are about encryption algorithms, since the top weighted words include ``\emph{chip}", ``\emph{nsa}", ``\emph{key}" and ``\emph{algorithm}". We further verified that our interpretation is correct by looking at the top weighted documents of the topics. This confirms that our model has potential for use in real world applications}.

\section{Conclusions and Future Work}
We presented a novel, query-driven topic model to help identify topics of interest in large datasets. Instead of asking experts to define keywords for these topics, we implemented a concept words extractor to automatically extract concept words and used the GPU model, incorporating word-filtering, to improve interpretability and performance. To distinguish between different contexts for the same concept, we further introduced a subtopic modeling procedure. The procedure can automatically infer all subtopics without having to determine the optimal number of subtopics beforehand. Experimental results on three benchmark datasets demonstrate the model's promise. In the future, we plan to evaluate our model's performance using real-world, qualitative analysis use cases.

\section*{Acknowledgments}
This work is partly funded by the EPSRC (grant no. EP/T017112/1, EP/V048597/1). ZF receives the PhD studentship jointly funded by the University of Warwick and China Scholarship Council. YH and RP are supported by Turing AI Fellowships funded by UK Research and Innovation (UKRI) (grant nos. EP/V020579/1 and EP/N510129/1 respectively). 
\clearpage

\bibliographystyle{acl_natbib}
\bibliography{acl2021}

\onecolumn
\newpage
\appendix
\section*{Appendix A: Query and keywords for each category}
\setcounter{table}{0}
\renewcommand{\thetable}{A\arabic{table}}
\begin{table*}[htb]
\centering
\resizebox{\textwidth}{!}{\begin{tabular}{lcccccccccc}
\toprule 
{\bf Concept Phrase} & \multicolumn{10}{c}{\bf Expert Defined Keywords} \\
\midrule 
{\bf atheism} & Agnosticism & theism & deism & islam & paganism & moral & atheist & religions & argument & exist \\
{\bf computer graphics} & image & digital & visual & 3d & 2d & visualization & print & geometry & synthesizing & processing \\
{\bf pc hardware} & cpu & monitor & keyboard & memory & card & sound & speakers & motherboard & power & pc \\
{\bf mac hardware} & touchpad & touchbar & drive & apple & mac & ram & gpu & system & sensors & physical \\
{\bf for sale} & product & mail & discount & bargain & shopping & price & sale & propertise & rent & summer \\
{\bf automobile} & car & vehicle & transportation & wheel & tire & road & parking & gasoline & energy & driver \\
{\bf motorcycles} & bike & scooters & mopads & motorbikes & trowel & commute & helmet & ride & speed & harley \\
{\bf baseball} & player & ball & small & hit & team & fielding & batting & runs & nbl & baseball \\
{\bf hockey} & puck & nhl & hockey & ice & rink & canada & rubber & curve & skater & guard \\
{\bf encrypt} & encoding & decryption & cryptographic & secure & plaintext & ciphertext & key & algorithm & pseudo & private \\
{\bf electronics}  & equipment & science & electricity & wire & console & computer & outlet & engineering & power & voltage \\
{\bf medicine} & medicine & surgery & hospital & climic & doctor & nurse & healthcare & symtoms & prescription & pharmacy \\
{\bf space} & rocket & nasa & astronomy & explore & moon & outerspace & spaceship & telescope & satellite & orbit \\
{\bf christian} & belief & faith & church & christianity & ethics & culture & ritual & Jesus & bible & truth \\
{\bf guns} & law & regulation & usa & victim & murder & violence & litigation & debate & firearms & legal \\
{\bf middle east} & israel & Iran & Iraq &  war & territory & turkey & attack & soldier & turkey & government \\
\bottomrule 
\end{tabular}}
\caption{\label{font-table} Concept phrases and expert defined keywords for the 20NewsGroup dataset.}
\end{table*}

\begin{table*}[htb]
\centering
\resizebox{1 \textwidth}{!}{\begin{tabular}{lcccccccccc}
\toprule 
{\bf Concept Phrase} & \multicolumn{10}{c}{\bf Expert Defined Keywords} \\
\midrule 
{\bf business} & bank & stock & market & business & economy & financial & investor & profit & price & deal \\
{\bf entertainment} & film & movie & music & tv & theater & festival & actor & show & book & hollywood \\
{\bf health} & drug & health & cancer & patient & disease & medical & hospital & healthcare & doctor & treatment \\
{\bf technology} & apple & google & sony & facebook & internet & mobile & ipad & technology & microsoft & phone \\
{\bf sport game} & league & win & player & team & tournament & game & playoff & sport & championship & point \\
\bottomrule
\end{tabular}}
\caption{\label{font-table} Concept phrases and expert defined keywords for the TagMyNews dataset.}
\end{table*}

\begin{table*}[htb]
\centering
\resizebox{1 \textwidth}{!}{\begin{tabular}{lcccccccccc}
\toprule 
{\bf Concept Phrase} & \multicolumn{10}{c}{\bf Expert Defined Keywords} \\
\midrule 
{\bf Business} & bank & stock & market & business & economy & financial & investor & profit & price & deal \\
{\bf Computers} & computer & software & programming & parallel & computing & memory & hardware & driver & cpu & processor \\
{\bf Culture Arts Entertainment} & movie & music & art & film & artist & museum & fashion & culture & imdb & actor \\
{\bf Education Science} & research & science & journal & university & student & education & scientific & mathematics & theory & school \\
{\bf Car Engineering} & engine & electrical & car & wheel & model & automobile & industrial & vehicle & cylinder & jet \\
{\bf Health} & drug & health & cancer & patient & disease & medical & hospital & healthcare & doctor & treatment \\
{\bf Politics Society} & political & party & democracy & government & republic & parliamentary & representative & president & communist & congress \\
{\bf Sports} & league & football & player & team & tournament & game & basketball & sport & hockey & championship \\
\bottomrule 
\end{tabular}}
\caption{\label{font-table} Concept phrases and expert defined keywords for the SearchSnippets dataset.}
\end{table*}

\clearpage
\section*{Appendix B: Topics generated for target concepts}
\setcounter{table}{0}
\renewcommand{\thetable}{B\arabic{table}}
\begin{table*}[h]
\begin{center}
\resizebox{\textwidth}{!}{\begin{tabular}{lcccccccccc}
\toprule 
{\bf Concept Phrase} & \multicolumn{10}{c}{\bf Top 10 words} \\
\midrule 
{\bf atheism} & people & question & thing & god & strong & point & make & belief & argument & evidence \\
{\bf computer graphics} & support & file & version & image & list & graphic & program & information & screen & address \\
{\bf pc hardware} & drive & pc & disk & scsi & software & modem & port & hard & controller & system \\
 
{\bf mac hardware} & card & mac & monitor & apple & system & video & problem & write & chip & work \\
 
{\bf for sale} & sell & sale & sold & original & interested & included & price & offer & box & cd \\
 
{\bf automobile} & car & auto & automobile & engine & ford & problem & mile & v6 & oil & dealer \\
 
{\bf motorcycles} & bike & riding & ride & motorcycle & rider & battery & buying & dog & back & dod \\
 
{\bf baseball} & run & baseball & game & pitcher & year & hit & player & average & team & good \\
 
{\bf hockey} & game & hockey & team & nhl & night & goal & player & coach & cup & year \\
 
{\bf encrypt} & key & message & chip & encryption & government & clipper & algorithm & system & phone & encrypted \\
 
{\bf electronics} & company & power & line & led & electronics & electronic & circuit & output & work & signal \\
 
{\bf medicine} & patient & medical & treatment & doctor & disease & study & clinical & medicine & food & effect \\
 
{\bf space} & space & nasa & shuttle & launch & satellite & station & moon & 1st & cost & orbit \\
 
{\bf christian} & god & christ & church & christian & love & word & bible & jesus & protestant & truth \\
 
{\bf guns} & gun & control & weapon & child & people & police & fire & law & handgun & amendment \\
 
{\bf middle east} & israel & arab & armenian & jew & israeli & muslim & people & middle & east & war \\
\bottomrule
\end{tabular}}
\end{center}
\caption{\label{font-table} Topics of target concepts for the 20NewsGroup dataset.}
\end{table*}

\begin{table*}[h]
\begin{center}
\resizebox{\textwidth}{!}{\begin{tabular}{lcccccccccc}
\toprule
{\bf Concept Phrase} & \multicolumn{10}{c}{\bf Top 10 words} \\
\midrule
{\bf Business} & business & management & marketing & trade & service & market & law & export & job & stock \\
 
{\bf Computers} & computer & computing & software & web & application & system & programming & apple & memory & chip \\
 
{\bf Culture Arts Entertainment} & art & culture & music & american & artist & tradition & history & ancient & museum & band \\
 
{\bf Education Science} & science & education & research & scientific & undergraduate & biology & journal & school & university & fiction \\
 
{\bf Car Engineering} & car & model & engineering & automobile & engine & wheel & auto & electrical & product & motor \\
 
{\bf Health} & health & disease & care & cancer & public & nutrition & information & medical & gov & drug \\
 
{\bf Politics Society} & political & party & democracy & system & military & politics & government & conflict & gov & war \\
 
{\bf Sports} & football & team & soccer & game & sport & hockey & news & tennis & score & player \\
\bottomrule
\end{tabular}}
\end{center}
\caption{\label{font-table} Topics of target concepts for the SearchSnippets dataset.}
\end{table*}

 
 
 
 

\clearpage
\section*{Appendix C: Algorithms}
\begin{algorithm}[htb]
\small
\SetAlgoLined
\ForEach {document $j \in D $}{ 
    $k \gets z\neq\tilde{z}_Q \sim Multinomial(1/K)$\;
    \ForEach {position $i \in j $}{
        \ForEach {$W_q \in W_Q $}{
            \If{$w_{ji} \in W_q$}{
                $k \gets \tilde{z}_q$\;
                break\;
            }
        }
    }
    $t \gets 0$\;
    \ForEach {position $i \in j $}{
        $S_{j,w_{ji}} \gets 0$\;
        $t_{ji} \gets t$\;
        $k_{jt} \gets k$\;
        UpdateCounter($S_{j,w_{ji}}$, $t_{ji}$, $k_{jt}$, $True$)\;
        $t \gets t+1$\;
    }
}
\caption{Initialize($K$, $D$, $W_Q$)}
\end{algorithm}

\begin{algorithm}[H]
\small
\SetAlgoLined
\eIf{$operation == True$}{
    \If{$t == t^{new}$}{
        $m_k \gets m_k + 1$\; 
    }
    \eIf{$S_{j,w} == 1$}{ 
        /* To apply GPU */\;
        \ForEach{$(w_i,w_q) \in \mathcal{M}$}{
            \If{$w == w_i$}{
                $n_{jt} \gets n_{jt}+A_{i,q}$\;
                $n_{kw} \gets n_{kw}+A_{i,q}$\;
            }
        }
    }{
        $n_{jt} \gets n_{jt} + 1$\;
        $n_{kw} \gets n_{kw} + 1$\;
    }
}{
    \eIf{$S_{j,w} == 1$}{
        /* To remove counts from GPU */\;
        \ForEach{$(w_i,w_q) \in \mathcal{M}$}{
            \If{$w == w_i$}{
                $n_{jt} \gets n_{jt}-A_{i,q}$\;
                $n_{kw} \gets n_{kw}-A_{i,q}$\;
            }
        }
    }{
        $n_{jt} \gets n_{jt} - 1$\;
        $n_{kw} \gets n_{kw} - 1$\;
    }
    \If{$n_{jt} == 0$}{
        $m_k \gets m_k - 1$\;
    }
}
\caption{UpdateCounter($S_{j,w}$, $t$, $k$, $operation$)}
\end{algorithm}

Note: $n_{kw}$ denotes the word count of $w$ in topic $k$; $\tilde{z}_Q$ denotes the concept related topics.
\end{document}